\documentclass{llncs}
\usepackage{makeidx}  % allows for indexgeneration
\usepackage{graphicx}
   \graphicspath{{./Figures/}}
   \DeclareGraphicsExtensions{.pdf,.eps,.svg,.png}
\usepackage{epsfig}
\usepackage{bbding}

\usepackage{booktabs}
\usepackage[caption=false]{subfig}
\usepackage{listings}
\usepackage{tabularx}

\usepackage{color}

\lstdefinelanguage{ptx}{
       basicstyle=\ttfamily,
      breaklines=true,
      morekeywords={ld,global,f32,cvt,s64,s32,shl,b64,add,}
}

\begin{document}
\frontmatter          % for the preliminaries
\pagestyle{headings}  % switches on printing of running heads
\addtocmark{Improving OpenCL Performance by Specializing Compiler Phase Selection and Ordering} % additional mark in the TOC

\mainmatter              % start of the contributions

\title{Improving OpenCL Performance by Specializing Compiler Phase Selection and Ordering}
\titlerunning{Improving OpenCL Performance by Specializing Compiler Phase Selection and Ordering}  % abbreviated title (for running head)
\author{Ricardo Nobre\Envelope%\inst{1,2}
\and Lu\'{i}s Reis%\inst{1,2}
\and Jo\~{a}o M. P. Cardoso%\inst{1,2}
}
\authorrunning{Ricardo Nobre et al.} % abbreviated author list (for running head)
\tocauthor{Ricardo Nobre, Lu\'{i}s Reis, and Jo\~{a}o M. P. Cardoso}
\institute{Faculty of Engineering, University of Porto\\
INESC TEC, Porto, Portugal\\
\email{rjfn@fe.up.pt},
\email{luis.cubal@fe.up.pt},
\email{jmpc@acm.org}
}

\maketitle              % typeset the title of the contribution

\begin{abstract}
Automatic compiler phase selection/ordering has traditionally been focused on CPUs and, to a lesser extent, FPGAs.
We present experiments regarding compiler phase ordering specialization of OpenCL kernels targeting a GPU.
We use iterative exploration to specialize LLVM phase orders on 15 OpenCL benchmarks to an NVIDIA GPU.
We analyze the generated NVIDIA PTX code for the various versions to identify the main causes of the most significant improvements and present results of a set of experiments that demonstrate the importance of using specific phase orders.
Using specialized compiler phase orders, we were able to achieve geometric mean improvements of $1.54\times$ (up to $5.48\times$) and $1.65\times$ (up to $5.7\times$) over PTX generated by the NVIDIA CUDA compiler from CUDA versions of the same kernels, and over execution of the OpenCL kernels compiled from source with the NVIDIA OpenCL driver, respectively.
We also evaluate the use of code-features in the OpenCL kernels. More specifically, we evaluate an approach that achieves geometric mean improvements of $1.49\times$ and $1.56\times$ over the same OpenCL baseline, by using the compiler sequences of the $1$ or $3$ most similar benchmarks, respectively.


\end{abstract}

\section{Introduction} \label{intro}

High Performance Computing (HPC) is increasingly relying on more heterogeneity, e.g., by combining Central Processing Units (CPUs) with accelerators in the form of Graphics Processing Units (GPUs) programmed with languages such as OpenCL~\cite{khronosopenclworkinggroup2015opencl} or CUDA~\cite{Nickolls:2008:SPP:1365490.1365500}.
For example, the Swiss National Supercomputing Centre's Piz Daint achieves a Linpack performance of 19.6 PFlop/s with a combination of Intel Xeon E5 V3 CPUs and NVIDIA P100 GPUs~\cite{pizdaint}.

Heterogeneous systems are used to achieve energy efficiency and/or performance levels that are not achievable on a single computing device/architecture.
For example, matrix multiplication is much faster on GPUs than CPUs for the same power/energy budget~\cite{gpuenergyefficiency}.
These accelerators offer a large number of specialized cores that CPUs can use to offload computation that exhibits data-parallelism and often other types of parallelism as well (e.g., task-level parallelism).
This adds an extra layer of complexity if one wants to target these systems efficiently, which in the case of HPC systems such as supercomputers is of utmost importance.
An inefficient use of the hardware is amplified by the magnitude of such systems (hundreds/thousands of CPU cores and accelerators), potentially resulting in large increases in the utilization/power bill and/or exacerbating cooling challenges as a consequence.
Ensuring that the hardware is efficiently used is in part the responsibility of the compiler used (e.g., GCC~\cite{gcc} and Clang/LLVM~\cite{clang,llvm}) to target the code to the computing devices in such systems and also the responsibility of the compiler users.
The programmer(s) and the compiler(s) have to be able to target different computing devices, e.g., CPU, GPU, and/or Field Programmable Gate Array (FPGA), and/or architectures, e.g., systems with ARM and x86 CPUs, in a manner that achieves suitable results for certain metrics, such as execution time and energy efficiency.

Compiler users tend to rely on the standard compiler optimization levels, typically represented by flags such as GCC's \texttt{-O2} or \texttt{-O3}, when using compilers such as GCC and Clang/LLVM.
These flags represent fixed sequences of analysis and transformation compiler passes, also referred to as compiler phase orders.
Programs compiled with these flags tend to outperform the unoptimized equivalent.
However, there are often other assembly/binary representations of the source application in the solution space with higher performance than the ones achieved through the use of the standard optimization levels~\cite{Kulkarni:2012:MCO:2384616.2384628,Purini:2013:FGO:2400682.2400715,Martins:2016:CSE:2899032.2883614,Nobre:2015:UPA:2764967.2764978,Nobre:2016:GIC:2907950.2907959}.
There are a number of scenarios that benefit from specialized compiler sequences, as there is potential to achieve considerable performance, energy or power improvements in comparison with what is achieved with the standard optimization levels.
Domains such as embedded systems or HPC tend to prioritize metrics such as energy efficiency that typically receive less attention from the compiler developers, so these domains can benefit further from these specialized sequences~\cite{Nobre:2016:CPC}.

Heterogeneous systems are composed of multiple devices with different architectures, each one needing different optimization strategies.
%For this reason, different optimization strategies are needed for each computing device.
With phase selection/ordering, one can achieve higher optimization for these devices, by specifying different compiler sequences for each of them.
In addition, the use of compiler phase selection/ordering specialization can reduce engineering costs, as
in a number of cases the same source code can be used when targeting architecturally different computing devices and/or different metrics through the use of different compiler phase orders.
This can reduce or mitigate the need to develop, manually optimize, and maintain multiple versions of the same function/application.

Automatic approaches have been proposed by a number of researchers in the context of compilation for CPUs and FPGAs.
These approaches can rely on iterative optimization algorithms such as sequential insertion~\cite{huangfpga}, simulated annealing (SA)~\cite{nobresa} and genetic algorithms (GAs)~\cite{Martins:2016:CSE:2899032.2883614}, where new compiler phase orders are iteratively generated and evaluated.
Other approaches rely on machine learning techniques (see, e.g.,~\cite{amir2016,amir,Sher:2014:PRN:2568326.2568328}), mapping a number of static and/or dynamic code features to compiler phase selections and/or phase orders.
 
This paper presents our work on identification of phase orderings when targeting OpenCL kernels to GPUs using LLVM.
The contributions of this paper are the following:
\begin{enumerate}
\item Assess the performance improvement that can be achieved when targeting OpenCL kernels to an NVIDIA GPU (Pascal architecture) using compiler pass phase ordering specialization with the LLVM compiler toolchain, in comparison with both the use of LLVM without the use of phase ordering specialization and the default OpenCL and CUDA kernel compilation strategies to NVIDIA GPUs.
\item Compare performance between OpenCL and CUDA kernels implementing the same freely available and representative benchmarks using recent NVIDIA drivers and CUDA toolchain, on an NVIDIA GPU with an up-to-date architecture.
\item Further motivate the importance of compiler phase ordering specialization with three experiments. These experiments demonstrate that compiler sequences that lead to performance improvements on some benchmarks do not necessarily do the same for others, and that the order of the passes on compiler sequences have a significant impact on performance.
\item Propose and evaluate a simple feature-based scheme to suggest phase orders that is able to achieve significant speedups with a low number of evaluations.
\end{enumerate}

This paper is an extended version of work published in~\cite{nobreheteropar}.
Contributions 3 and 4 are content exclusive to this paper.

The rest of this paper is organized as follows.
Section~\ref{section-experiments} describes the methodology for the experiments presented in this paper.
Section~\ref{section:gpu_results} presents experiments performed to assess the performance improvement that can be achieved with phase ordering specialization when targeting OpenCL kernels to a NVIDIA GPU (Pascal architecture).
This section includes a set of additional experiments, targeting the NVIDIA GPU, performed to further motivate the use of efficient compiler phase ordering specialization approaches.
Finally, it also includes an explanation at the NVIDIA PTX assembly level (for OpenCL compilation with LLVM and for CUDA compilation with the NVIDIA CUDA Compiler (NVCC)), for each kernel, of the causes of the performance improvements achieved with specialized compiler sequences.
Section~\ref{section:features} presents results, targeting the same NVIDIA GPU, for the use of a simple approach that relies on OpenCL code features to suggest the use of specific compiler sequences for the compilation of a new unseen OpenCL kernel.
Section~\ref{section:relatedwork} presents related work in the context of compiler phase ordering.
Final remarks about the presented work and ongoing and future work are presented in Section \ref{section-conclusions}.

\section{Experimental Setup} \label{section-experiments}

We extended our compiler phase selection/ordering Design Space Exploration (DSE) system, previously used in~\cite{Nobre:2016:GIC:2907950.2907959,Nobre:2016:CPC},
to support exploring compiler sequences targeting NVIDIA GPUs using Clang/LLVM 3.9~\cite{clang} and the libclc OpenCL library 0.2.0~\cite{libclc}.

\subsection{Target systems and system configurations}
 
The NVIDIA GPU used for the experiments is a variant of the NVIDIA GP104 GPU in the form of an EVGA NVIDIA GeForce GTX 1070~\cite{gtx1070} graphics card (08G-P4-6276-KR) with a 1607/1797 MHz base/boost graphics clock and 8GB of 256 bit GDDR5 memory with a transfer rate of 8008 MHz (256.3 GB/s memory bandwidth).
The GPU is connected to a PCI-Express 3.0 16x interface on its respective workstation.
For the experiments targeting a NVIDIA GPU, we used a workstation with an Intel Xeon E5-1650 v4 CPU, running at 3.6 GHz (4.0 GHz Turbo) and 64 GB of Quad-channel ECC DDR4 @2133 MHz.
We relied on Ubuntu 16.04 64-bit with the NVIDIA CUDA 8.0 toolchain (released in Sept. 28, 2016) and the NVIDIA 378.13 Linux Display Driver (released in Feb. 14, 2017).
The NVIDIA GPU is set to persistence mode with the command \texttt{nvidia-smi -i <target gpu> -pm ENABLED}. % tava ENABLE (estava mal)
This forces the kernel mode driver to keep the GPU initialized at all instances, avoiding the overhead caused by triggering GPU initialization at application start.
The preferred performance mode is set to \emph{Prefer Maximum Performance} under the \emph{PowerMizer settings} tab in the \emph{NVIDIA X Server Settings}, in order to reduce the occurrence of extreme GPU and memory frequency variation during execution of the GPU kernels.

In order to reduce DSE overhead, and given the fact that we found experimentally that multiple executions of the same compiled kernel had a small standard deviation in respect to measured wall time, each generated code is only executed a single time during DSE.
Only in a final phase on the DSE process are the top solutions executed $30$ times and averaged in order to select a single compiler phase order.
All execution time metrics reported (baseline CUDA/OpenCL and OpenCL optimized with phase ordering) in this paper correspond to the average over $30$ executions.

\subsection{Kernels and objective metric}

In this paper, we use kernels from the PolyBench/GPU benchmark~\cite{polybenchgpu} suite to assess the potential for improvement with phase ordering specialization when targeting a NVIDIA GPU.
This benchmark suite includes kernels from 15 benchmarks from different domains which represent computations that would be performed on GPUs in the context of HPC, including convolution kernels (\texttt{2DCONV}, \texttt{3DCONV}), linear algebra (\texttt{2MM}, \texttt{3MM}, \texttt{ATAX}, \texttt{BICG}, \texttt{GEMM}, \texttt{GESUMMV}, \texttt{GRAMSCH}, \texttt{MVT}, \texttt{SYR2K}, \texttt{SYRK}), data mining (\texttt{CORR}, \texttt{COVAR}), and stencil computations (\texttt{FDTD-2D}).
PolyBench/GPU includes implementations of these kernels in CUDA, OpenCL, and HMPP. 
Finally, this benchmark suite is freely available and thus contributes for making the results presented in this paper reproducible.

We experimented with both the CUDA and the OpenCL implementations available for each PolyBench/GPU benchmark.
We rely on the default dataset shape so that reproducibility of the performance metrics reported in this paper is more straightforward.

We performed the minimum of changes to ensure a fair comparison between the OpenCL and the CUDA versions.
This included modifying the benchmarks to ensure that both use the same floating-point precision.
In addition, we also modified the initialization/validation code of some kernels to detect errors produced by some phase orders, by initializing data to non-zero values and adding missing checks.

\subsection{Compilation and execution flow with specialized phase ordering}

\begin{figure}[t]
\centering
\includegraphics[scale=0.4]{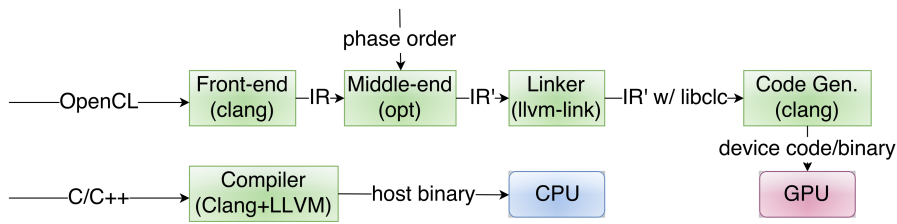}
\caption{Compilation targeting a host CPU and an OpenCL-compatible GPU using Clang/LLVM and libclc. Phase orders are generated/selected by our DSE system.}
\label{fig:compilationflow}
\end{figure}

Figure~\ref{fig:compilationflow} depicts the compilation flow we use for targeting the host CPU and the GPU, with the use of specialized phase orders when compiling for the latter.
We use Clang compiler's OpenCL frontend with the libclc library to generate an LLVM assembly representation of a given input OpenCL kernel.
The libclc library is an open source library with support for OpenCL functions on AMD and NVIDIA GPUs. % that implements functions as specified in OpenCL 1.1.

We use the LLVM Optimizer tool (\texttt{opt})%~\cite{llvmopt} 
to optimize the ~LLVM intermediate representation (IR) using a specific optimization strategy represented by a compiler phase order, and we link this optimized IR with the libclc OpenCL functions for our target using \texttt{llvm-link}.
Finally, using Clang, we generate the NVIDIA PTX~\cite{nvidiaptx} from the LLVM bytecode resulting from the previous step, using the \texttt{nvptx64-nvidia-nvcl} target.
PTX is NVIDIA's IR for GPU computations, and is used by NVIDIA's OpenCL and CUDA implementations.
Although PTX is not the final assembly executed on the GPU, it is typically the closest representation one is able to access without direct access to the internals of NVIDIA's drivers.

In the code that executes in the host, each PTX representation generated from the optimized LLVM IR is used to create an OpenCL program object instead of loading the OpenCL kernel source file that comes with the specific PolyBench/GPU benchmark.
To do this, we used offline compilation. That is, we load the compiled NVIDIA PTX code with the \texttt{clCreateProgramWithBinary} function instead of passing the OpenCL source code to \texttt{clCreateProgramWithSource}, which is the most commonly used mechanism in OpenCL.

A compiler phase order represents not only the compiler passes to execute in the compiler optimizer, which can be in the order of the hundreds, but also their order of execution. For instance, as of LLVM 3.9, the phase order for --O3 includes the execution of a total of 172 instances of compiler passes, from which only 72 represent the execution of distinct passes. %For reference, the \texttt{opt} --O2 sequence is the same as \texttt{opt} --O3, except for not having the \texttt{-argpromotion} pass.
The fact that compiler passes are interdependent and interfere with each other's execution in ways that are difficult to predict can make it extremely hard to manually generate suitable compiler sequences.

\subsection{Validation of the code generated after phase ordering}

Each PolyBench/GPU benchmark has verification embedded in its code that consists in executing the OpenCL GPU kernel(s) followed by the execution of a functionally equivalent sequential C version on the CPU, and comparing the two.
This alone poses a challenge, as CPU execution using the same parameters as the ones used for GPU execution takes a long time for a considerable number of PolyBench/GPU benchmarks.
This would have an unreasonable impact on the phase ordering exploration time.

During DSE, we separate validation from the measurement phases to reduce the time for each DSE iteration.
We validate the code generated by compilation with each particular compiler phase order by executing the unoptimized serial version on the CPU (as in the original PolyBench/GPU code) and the optimized version on the GPU with inputs that can be processed quickly.
However, we also execute the same GPU code using the original inputs in order to measure the execution time.

We further reduce exploration time by checking whether an identical optimized version (i.e., NVIDIA PTX) was previously generated. If so, we reuse the results (i.e., correctness and measured performance) from that previous execution.

At the end of phase ordering exploration, all compiler pass sequences that were iteratively evaluated during DSE are ordered by their resulting fitness regarding the objective metric of interest.
With performance as objective metric, sequence/fitness pairs are ordered from the one resulting in fastest execution time to the one resulting in least performance.
Then, as a final step, the optimized version that resulted in highest performance is executed with the original inputs on both the non-optimized CPU version and the optimized GPU version, with $30$ randomly generated inputs that result in the same number of each kind of GPU instructions being executed.
We choose the fastest version that passes validation.

This is performed to eliminate possible situations where a compiled OpenCL code gives correct results using a small input set but gives wrong results with the original input set.
This is just a precaution, and in our tests we did not encounter any case where this was a problem.

The PolyBench/GPU kernels are mostly composed of floating-point operations and the result of floating-point operations can be affected by reordering operations and rounding.
Because of this, we allow for up to 1\% difference between the outputs of CPU and GPU executions when checking if a given compiler phase order results in code that generates valid output.

\section{Impact of Phase Selection and Ordering} \label{section:gpu_results}

We present in this section results of experiments consisting in compiling/evaluating the OpenCL kernels from each of the PolyBench/GPU benchmarks with a set of $10,000$ randomly generated compiler phase orders (given a compiler/target pair, the same set of phase orders was used with all OpenCL codes) composed of up to $256$ pass instances.
We allow any given sequence of LLVM passes to include repeated calls to the same pass.
Passes were selected from a list with all LLVM passes, except passes with names starting with \texttt{-view-*} and the ones that resulted in compilation and/or execution problems when used individually to compile the PolyBench/GPU OpenCL kernels.

For each of the benchmarks, we measured the execution times for the CUDA version, the original OpenCL version compiled from source, an offline compiled OpenCL without optimization, an offline compiled OpenCL with standard LLVM optimization levels (i.e., the best of \texttt{-O1}, \texttt{-O2}, \texttt{-O3} and \texttt{-Os} for each benchmark, which we will refer to as \texttt{-OX}) and an offline compiled OpenCL with our custom compiler optimization phase orders.

\subsection{Performance evaluation}

We compared the results for the various versions of the benchmarks (offline OpenCL versions, OpenCL from source and CUDA) to determine how they perform. Using custom phase orders found by iterative compilation produced code that consistently outperforms the other OpenCL variants, and nearly always outperforms the CUDA version.

\begin{figure}[t]
\centering
\includegraphics[scale=0.69]{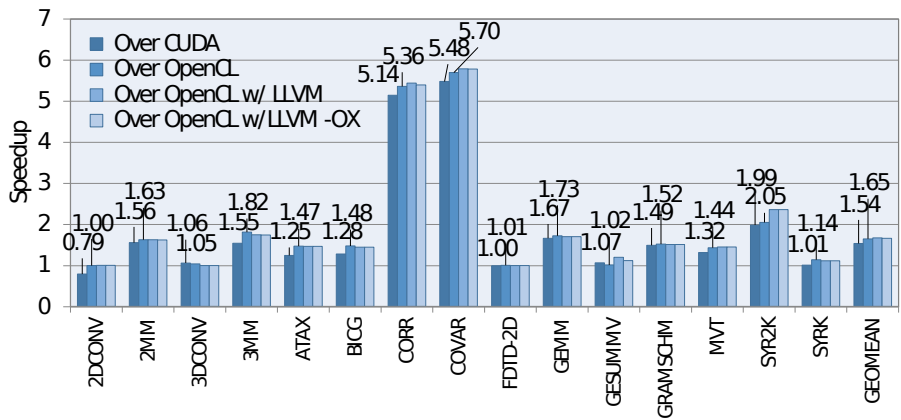}
\caption{Performance improvements from phase ordering with LLVM over CUDA (\emph{Over CUDA}) and OpenCL using the default compilation method (\emph{Over OpenCL}) for the NVIDIA GP104 GPU and over OpenCL to PTX compilation using Clang/LLVM without (\emph{Over OpenCL w/LLVM}) and with the standard optimization levels (\emph{Over OpenCL w/LLVM --OX}). Data labels are shown for the CUDA and OpenCL using the default compilation method.}
\label{fig:phase_ordering_gtx1070}
\end{figure}

Figure \ref{fig:phase_ordering_gtx1070} depicts the speedups obtained with phase ordering over the standard OpenCL (compiled from source), CUDA and the other OpenCL (offline compilation with LLVM) baselines.
With phase ordering specialization we were able to achieve a geometric mean speedup of 1.54$\times$ over the CUDA version and a speedup of 1.65$\times$ over the execution of the OpenCL kernels compiled from source.
Additionally, code compiled with specialized phase ordering can be up to 5.48$\times$ and up to 5.7$\times$ faster than the respective CUDA implementation and the OpenCL compiled from source.
Table \ref{tab_phase_orders} depicts the LLVM 3.9 compiler phase orders that result in the performance improvement factors presented in Figure~\ref{fig:phase_ordering_gtx1070}.

\begin{table}[t]
\caption{LLVM 3.9 compiler phase orders that resulted in compiled kernels with highest performance when targeting an NVIDIA GP104 GPU. Compiler passes that resulted in no performance improvement were eliminated from the compiler phase orders. No compiler phase orders resulted in improving the performance of 2DCONV, 3DCONV or FDTD-2D, when targeting the NVIDIA GPU.}
\label{tab_phase_orders}
\begin{tabularx}{\textwidth}{ l X }
\toprule
 \textbf{Benchmark} & \textbf{Compiler Phase Order} \\
\midrule
% 2DCONV 	& \emph{Empty Phase Order} \\
 2MM 	& \texttt{-cfl-anders-aa -dse -loop-reduce -licm -instcombine} \\
% 3DCONV 	& \emph{Empty Phase Order} \\
 3MM 	& \texttt{-loop-reduce -gvn-hoist -reg2mem -cfl-anders-aa -sroa -licm} \\ 
 ATAX 	& \texttt{-bb-vectorize -loop-reduce -licm -cfl-anders-aa} \\
 BICG 	& \texttt{-gvn -loop-reduce -cfl-anders-aa -licm -loop-reduce} \\
 CORR 	& \texttt{-cfl-anders-aa -loop-reduce -gvn -sink -loop-extract-single -loop-unswitch -loop-unswitch -ipsccp -reg2mem -licm -nvptx-lower-alloca} \\
 COVAR 	& \texttt{-cfl-anders-aa -loop-unswitch -reassociate -jump-threading -loop-reduce -gvn -loop-unswitch -reassociate -sink -loop-unswitch -loop-reduce -jump-threading -reg2mem -licm -nvptx-lower-alloca} \\
% FDTD-2D 	& \emph{Empty Phase Order} \\
 GEMM 	& \texttt{-cfl-anders-aa -print-memdeps -loop-reduce -licm} \\
 GESUMMV 	& \texttt{-instcombine -reg2mem -mem2reg} \\
 GRAMSCHM 	& \texttt{-sink -reg2mem -licm -cfl-anders-aa -sroa} \\
 MVT 	& \texttt{-gvn -loop-reduce -cfl-anders-aa -licm} \\
 SYR2K 	& \texttt{-loop-reduce -loop-unroll -instcombine -loop-reduce -licm -cfl-anders-aa} \\
 SYRK	& \texttt{-licm -cfl-anders-aa -reg2mem -licm -sroa} \\
\bottomrule
\end{tabularx}
\end{table}

There were no significant performance difference between the offline compilation model using Clang/LLVM without custom phase ordering and the OpenCL versions from source.
Likewise, using the LLVM standard optimization level flags did not result in noticeable improvements in terms of the performance of the generated code for most benchmarks.
For \texttt{2DCONV}, \texttt{3DCONV}, \texttt{FDTD-2D} and \texttt{SYR2K} none of the standard optimization level flags resulted in different code being generated.
For benchmarks \texttt{2MM}, \texttt{3DCONV}, \texttt{3MM}, \texttt{ATAX}, \texttt{BICG}, \texttt{GEMM}, \texttt{GESUMMV}, \texttt{GRAMSCHM}, \texttt{MVT} and \texttt{SYRK}, the generated code using the optimization level flags differs from the generated code without optimizations. However, the different optimization levels (i.e., \texttt{-O1}, \texttt{-O2}, \texttt{-O3}, \texttt{-Os}) all produced the same code.

\texttt{CORR} and \texttt{COVAR} are the only benchmarks for which different optimization level flags produce different code. However, even in these benchmarks, the performance impact was usually minimal (within 1\%).
The only exceptions were \texttt{GESUMMV} and \texttt{GRAMSCHM}.
In the case of \texttt{GESUMMV}, the use of all tested optimization levels resulted in 1.07$\times$ performance improvement over the non-optimized version.
For \texttt{GRAMSCHM}, the non-optimized version was 1.04$\times$ faster than all the versions produced by the optimization level flags.

The difference between the OpenCL baselines is that one represents the de facto OpenCL compilation flow (with OpenCL program created from source) and the others represent the compilation using LLVM (with OpenCL program created from binary) using the standard optimization level that results in the generation of code with highest performance on a kernel-by-kernel basis, and compilation using LLVM but with no optimization.
Finally, on these benchmarks, performance with CUDA tends to be better than with OpenCL, if no specialized phase ordering is considered.
The geometric mean (considering all 15 PolyBench/GPU benchmarks) of the performance improvement with CUDA (over OpenCL from source) is 1.07$\times$.
The \texttt{2DCONV}, \texttt{3MM}, \texttt{ATAX}, \texttt{BICG} and \texttt{SYRK} benchmarks are at least 1.1$\times$ faster in CUDA than in OpenCL.
All other benchmarks with exception for \texttt{3DCONV} and \texttt{GESUMMV} are still faster in CUDA than in OpenCL, although by a smaller margin.

We also evaluated the same PolyBench/GPU kernels with compiler phase ordering using LLVM when targeting an AMD Fiji GPU using the ROCm 1.6 drivers, compared with the code generated by online compilation and with the use of LLVM without phase ordering, and were able to achieve speedups both over compilation from source ($1.65\times$) and compilation with LLVM using the standard optimization levels ($1.73\times$).
These results can be of special interest, as the compilation approach for AMD and NVIDIA are significantly different.
When targeting AMD GPUs, LLVM generates the final AMD GPU ISA code, whereas the LLVM backend targeting NVIDIA GPUs generates NVIDIA's PTX, which is an intermediate representation that is further optimized by the CUDA compiler.

The speedups using phase ordering differ substantially from the ones obtained when targeting the NVIDIA GPU, with benchmarks \texttt{2MM}, \texttt{3DCONV}, \texttt{3MM}, \texttt{GEMM}, \texttt{GESUMMV}, \texttt{GRAMSCHM} and \texttt{SYRK} improving substantially more with phase ordering specialization when targeting the AMD Fiji GPU, while \texttt{CORR}, \texttt{COVAR}, \texttt{MVT} and \texttt{SYR2K} benefited more when targeting the NVIDIA GPU.
This reinforces the notion that the efficiency of custom phase orders is highly device-dependent, so target specialization is important.

\subsection{Problematic phase orders}

Considering the evaluation of all $10,000$ sequences with all PolyBench/GPU benchmarks (\texttt{TOTAL} column), the most common problem is the report being non-existent or broken ($17\%$), the second is the generation of incorrect output by the compiled OpenCL kernels ($13\%$), and the third (and last) is the optimized LLVM IR not being generated ($3\%$).

Typically, the non-existence of optimized LLVM IR after calling the LLVM Optimizer is caused by a compiler crash.
In some cases, the execution of the optimized/compiled kernels does not terminate.
This can happen because of problems in the kernel themselves or because the compiled kernels are not given enough time to finish execution.
Our phase ordering exploration system has a timeout parameter for limiting the overhead of exploration allowed to the execution of the OpenCL kernels compiled after phase ordering.
The fact there is a timeout is not detrimental in the sense that it does not result in discarding any given suitable compiler phase orders in the context of performance maximization, because if a compiled OpenCL kernel takes too much time to execute, it means that it was not compiled with a phase order suitable to maximize performance.
Although compiler developers typically make an effort to assure that any given compiler pass that transforms an IR will have as output other IR that is functionally equivalent to the original, this is difficult to implement in practice,
especially given the fact that compilers can have tens or hundreds of compiler passes and it is difficult to predict all possible interactions between passes.
For instance, Eide and Regehr evaluated thirteen production-quality C compilers and, for each, were able to find cases where incorrect code to access volatile variables was generated~\cite{eide}.
Moreover, given the fact that during our phase ordering exploration we are compiling with sequences that were possibly never tested/evaluated by the compiler developers, there is a greater potential for generating code that does not conform with the same functionality of the original code and that will generate outputs that are different than expected.

\subsection{Additional experiments}

We performed the following experiments in order to better assess the importance of specialization of compiler sequences when using LLVM to target NVIDIA GPUs:
\begin{enumerate}
\item Evaluate the use of the sequences found for any given PolyBench/GPU OpenCL benchmark (see Table~\ref{tab_phase_orders}) in the remaining benchmarks.
\item Show how the performance of the different PolyBench/GPU OpenCL benchmarks is impacted by different compiler sequences, comparing the performance differences registered for different kernels using the same set of compiler sequences.
\item Show the effect of the different phase orders, on each benchmark, constructed with permutations of the sequences found for the same benchmark (see Table~\ref{tab_phase_orders}).
\end{enumerate}

\begin{figure}[t]
\centering
\includegraphics[scale=0.65]{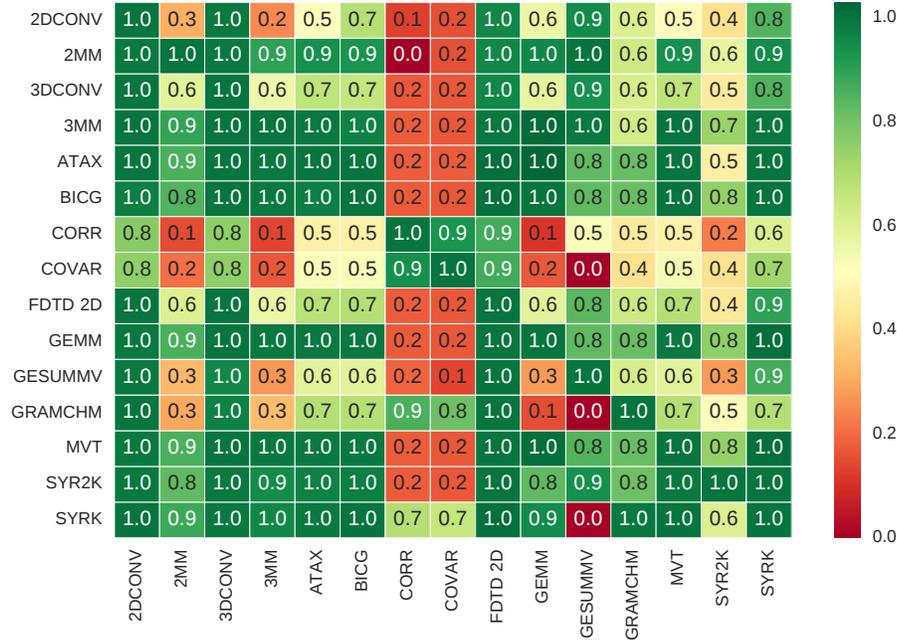}
\caption{Performance ratios for using sequences found for each of the benchmark in all benchmarks. The $X$ axis represents the benchmarks and the $Y$ axis represents the sequences (i.e., the best found for each benchmark). Performance ratios are represented with a precision of $5\%$, so values represented as $1.0$ can be as low as $0.95$. Values represented as $1.0$ but that are closer to $0.95$ are represented with a slightly lighter shade.
}
\label{fig:sequences_in_other_kernels}
\end{figure}

Figure~\ref{fig:sequences_in_other_kernels} shows the matrix resulting from evaluating the sequences from Table~\ref{tab_phase_orders} for compilation of each of the PolyBench/GPU OpenCL benchmarks.
Values, between $0$ and $1$, represent performance factors resulting from comparing with the performance obtained with the sequence found to be the best for each benchmark.

The phase orders found individually for each benchmark result in a very wide performance factor range when used to compile the OpenCL code of the remaining $14$ benchmarks.
Moreover, some benchmark/sequence pairs (\texttt{CORR}/\texttt{2MM}, \texttt{GESUMMV}/\texttt{COVAR}, \texttt{GESUMMV}/\texttt{GRAMSCHM}, \texttt{GESUMMV}/\texttt{SYRK}) did not pass validation, resulting in incorrect outputs when executing the generated code.

\begin{figure}[t]
\centering
\includegraphics[scale=0.605]{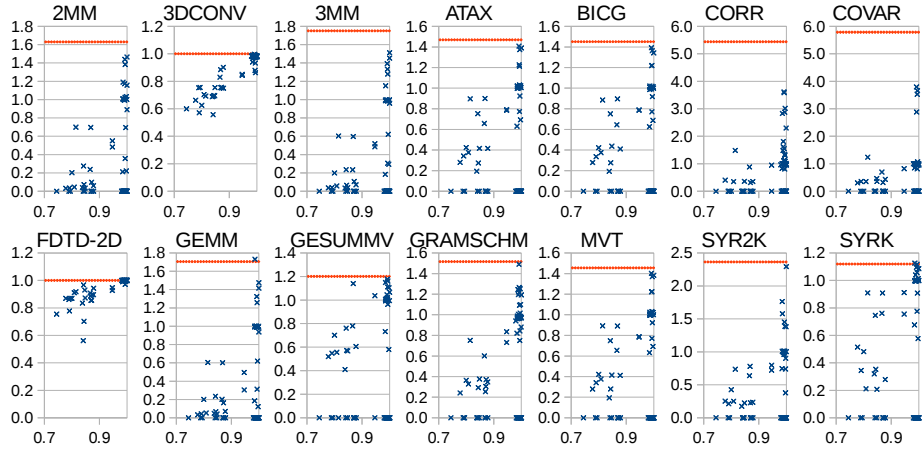}
\caption{Speedup for the same sequences on different kernels. Each point represents a phase order, with its $X$ and $Y$ position representing the speedup for the 2DCONV benchmark and for the benchmark indicated on top of each chart, respectively.
The horizontal reference line represents the speedup for the best found sequence for each benchmark.
The same baseline was used for all benchmarks so that comparisons between charts possible.}
\label{fig:100_seqs}
\end{figure}

Each individual compiler sequence has a different impact on different OpenCL kernels.
Figure \ref{fig:100_seqs} presents the performance impact (i.e., speedup) of the first $100$ compiler sequences evaluated during the initial DSE process (from the set of $10,000$ sequences evaluated) on each of the OpenCL benchmarks.
The performance baseline is offline compilation with LLVM with no optimization, as it was experimentally determined that the use of the standard optimization levels in LLVM only very rarely improves the performance of the generated NVIDIA PTX assembly code (see Figure \ref{fig:phase_ordering_gtx1070}).
The performance of the best phase order (see Table \ref{tab_phase_orders}) found for each OpenCL benchmark is represented as a horizontal reference line.
Speedups above this reference line are entirely caused by random performance variations.
For the \texttt{2MM}, \texttt{3MM}, \texttt{CORR} and \texttt{COVAR} OpenCL benchmarks, the first $100$ DSE iterations could not achieve the speedup achieved with the $10,000$ iterations.
It is also interesting to notice that there is a cluster of points from the compiler sequence solution space close to the baseline (i.e., close to horizontal line $y=1$), and that for some kernels the concentration of points close to the highest achieved speedup (i.e., close to the reference line) is very scarce (e.g., \texttt{GEMM}, \texttt{GRAMSCHM}, \texttt{SYR2K}), while for other kernels there are several sequences close to the best found (e.g., \texttt{SYRK}).
Benchmark/sequence pairs that result in generation of incorrect output, compiler crashes or execution timeout expiration, are represented on top of the $X$ axis (i.e., $y=0$).
These results show that: 1) given a compiler sequence randomly selected from the compiler sequence's space, it is most likely that it will generate code without performance improvements in relation to the code achieved without compiler phase ordering specialization; and 2) for some kernels, it is considerably less likely that a given compiler sequence will result in generated code that performs close to the performance achieved with a considerable number of iterations of iterative compilation.

\begin{figure}[t]
\centering
\includegraphics[scale=0.62]{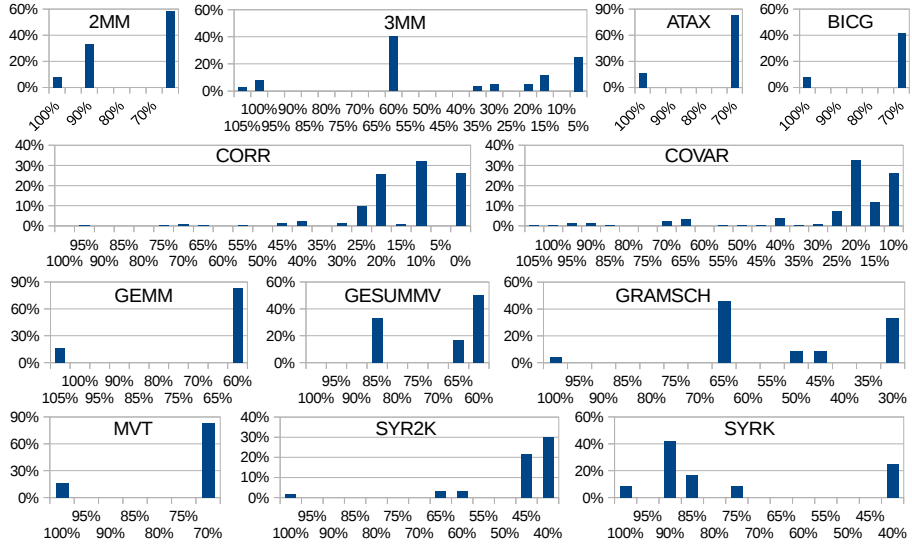}
\caption{Impact of the order of compiler passes in the best-found sequence for each benchmark. This is demonstrated in the form of distribution of speedups over the best order ($X$ axis). The $Y$ axis indicates the percentage of permutations that achieve that range of speedups.}
\label{fig:permutations}
\end{figure}

Figure~\ref{fig:permutations} shows the results obtained for experiments where permutations of the phase orders from Table~\ref{tab_phase_orders} are evaluated. 
Up to $1,000$ randomly generated permutations were evaluated for compilation of each kernel.
Each permutation includes all the compiler passes that are present in the phase order from Table~\ref{tab_phase_orders} (including the number of pass instances that are repeated in the sequence).
The \texttt{2DCONV}, \texttt{3DCONV} and \texttt{FDTD-2D} OpenCL benchmarks are not included because the initial DSE process could not find a LLVM phase order that resulted in improved performance.
The execution performance after compilation with a large number of permutations resulted in considerable performance degradation. 
Some of these permutations were only able to achieve $10\%$ or less of the execution performance achieved with the initial phase order (e.g., \texttt{3MM}, \texttt{CORR}, \texttt{COVAR}).
We believe these results strongly motivate, at least under some circumstances, the use of compiler phase ordering when targeting an NVIDIA GPU. 

\subsection{Insights about performance improvements}

In this subsection, we explain, for each PolyBench/GPU benchmark, some of the key reasons behind the performance improvement achieved with phase ordering, compared with the performance achieved by the OpenCL baselines and the baseline CUDA versions.
More specifically, we compare the PTX output resulting from OpenCL offline compilation with specialized phase ordering with PTX generated from OpenCL offline compilation without phase ordering and with PTX generated from the CUDA versions.

For \texttt{2DCONV}, CUDA is $1.26\times$ faster than all versions compiled from OpenCL. 
The compiler pass phase ordering DSE process was not able to find an LLVM sequence capable of optimizing this benchmark.
The main improvement regarding the CUDA version over the OpenCL versions seems to be the generation of more efficient code for loads from global memory. Figure~\ref{fig:ptx-load-comparison} shows the difference between the two approaches. Whereas CUDA load operations typically result in a single PTX instruction, the equivalent for OpenCL typically results in 5 PTX instructions. We believe that this difference in load instructions is the primary reason for CUDA's advantage over OpenCL for the \texttt{2DCONV} OpenCL code.

\begin{figure}[t]
  \centering
  \begin{minipage}{.45\textwidth}
   \scriptsize 
  \begin{lstlisting}[language=ptx]
  
  
  
 
ld.global.f32 	%f2, [%rd6+4]
  \end{lstlisting}
  \textbf{(a)} PTX load code generated from CUDA.
    \end{minipage}
  \centering
    \begin{minipage}{.45\textwidth}
     \scriptsize 
  \begin{lstlisting}[language=ptx]
add.s32 	%r17, %r14, %r1;
cvt.s64.s32	%rd16, %r17;
shl.b64 	%rd17, %rd16, 2;
add.s64 	%rd18, %rd1, %rd17;
ld.global.f32 	%f2, [%rd18];
  \end{lstlisting}
  \textbf{(b)} PTX load code generated from OpenCL (\texttt{-O3}). 
      \end{minipage}
\caption{PTX code for equivalent load operations, for CUDA and OpenCL (2DCONV benchmark).}
\label{fig:ptx-load-comparison}
\end{figure}

For \texttt{2MM}, the OpenCL version optimized with phase ordering is $1.63\times$ and $1.56\times$ faster than the OpenCL (compiled from source) and CUDA baselines, respectively. 
The main reason for these speedups is the removal of store operations within the kernel loop.
Both the OpenCL and the CUDA baseline versions of this kernel repeatedly overwrite the same element, to the detriment of program performance.
The phase ordered version instead uses an accumulator register and performs the store only after all the loop computations are complete, which substantially reduces the number of costly memory accesses.
It is unclear why the baseline OpenCL and CUDA versions do not perform this optimization.
One possibility is that the NVIDIA OpenCL/CUDA compiler and LLVM w/o the use of special phase orders are unable to determine that there are no aliasing issues.
In the context of the \texttt{2MM} benchmark, it is correct to assume that there is no aliasing, as any aliasing would result in a data race (in OpenCL 2.0), which is undefined behavior~\cite{khronosopenclworkinggroup2015opencl}.
We do not know if the optimization was applied because LLVM correctly discovered this fact, or if there is a bug that happened to result in correct code by accident.
Even if the optimization turns out to be the result of a bug, we believe this speedup represents an opportunity for approaches based on \emph{Loop Versioning} transformations~\cite{loopversioning}.
Although this benchmark uses two kernels, both are equivalent (the only difference being kernel and variable names), and thus the same analysis applies to both.
There are two differences between the CUDA version and OpenCL versions compiled offline that can explain the different execution times.
The first being the aforementioned issue with load instructions (see Figure~\ref{fig:ptx-load-comparison}), the second being a different loop unroll factor as the phase ordered version based on OpenCL uses efficient load instructions, but uses a loop unroll factor of $2$ (while the CUDA version uses an unrolling factor of $8$).

For \texttt{3DCONV}, we were unable to achieve a speedup using any of the compiler phase orders evaluated, when compared with LLVM w/ or w/o the optimization level flags.
We believe this happens because most of the time spent on the benchmark is due to global memory loads that are not removed or improved by any LLVM pass.
Any optimization will only modify the rest of the code, which takes a negligible amount of time compared to the memory operations.
Interestingly, there is a speedup from the use of the LLVM PTX backend compared with the OpenCL from source compilation path ($1.05\times$) and the compilation from CUDA ($1.06\times$). 

On the \texttt{3MM} benchmark, we were able to achieve speedups of $1.55\times$ and $1.82\times$ over the baseline CUDA and OpenCL version compiled from source, respectively. The main reason for the performance improvement is the removal of the memory store operation from the computation loop.

The OpenCL version of \texttt{ATAX} optimized with phase ordering achieves a speedup of $1.47\times$ and $1.25\times$ over the baseline OpenCL and CUDA versions, respectively.
Once again, the phase ordered version is able to move memory stores out of the innermost loops of the kernels, which explains the speedups.
Additionally, the difference between the CUDA and the baseline OpenCL versions can be explained by a different loop unroll factor ($2$ for OpenCL, $8$ for CUDA).
The CUDA version uses the previously described simpler code pattern for memory loads compared to the baseline OpenCL version, but the phase ordering version also uses an efficient memory load pattern.

On the \texttt{BICG} benchmark, we achieved a speedup of $1.48\times$ over OpenCL compiled from source, and $1.28\times$ over CUDA. Main differences between the versions are the memory stores in the kernel loop, the unroll factor and the inefficient memory access patterns in the baseline OpenCL version.

The \texttt{CORR} benchmark is one of the benchmarks that benefit the most from phase ordering ($5.36\times$ and $5.14\times$ over baseline OpenCL from source and CUDA versions, respectively).
The version generated by phase ordering contains several memory accesses to a local memory storage buffer (named \verb|__local_depot|) that serves a purpose similar to the stack of the CPU.
We believe this buffer is inserted due to the use of the \verb|reg2mem| pass without a corresponding \verb|mem2reg|.
These instructions do not seem to have a significant impact, either because they are eliminated in the compilation of the PTX to device-specific code by the NVIDIA GPU driver, or because the accesses to local memory are too fast to affect non-negligible performance variations.
Phase ordering is also capable of moving global memory stores out of loops, which neither the CUDA version nor the baseline OpenCL versions do.
In general, for this benchmark, the CUDA version tends to produce more compact load instructions and use higher loop unroll factors than the OpenCL versions.

The \texttt{COVAR} and \texttt{CORR} benchmarks both rely on the same \verb|mean_kernel| and \verb|reduce_kernel| functions. However, these functions represent only a fragment of the total execution code.
Regardless, the same conclusions from \texttt{CORR} apply to \texttt{COVAR}: phase ordering removes global stores from the loop, but introduces several new registers and local memory accesses.

The functions of the \texttt{FDTD-2D} benchmark are very straightforward, with little potential for optimization. As such, phase ordering had no impact. 

The performance differences for the \texttt{GEMM} benchmark ($1.67\times$ and $1.73\times$ over the OpenCL from source and the CUDA baselines) can be explained by the removal of the memory store operation from the kernel loop, the different unroll factor and the different pattern of memory load instructions.

There was only a small performance improvement for the \texttt{GESUMMV} benchmark ($1.07\times$ over CUDA and $1.02\times$ over the baseline OpenCL from source).
Phase ordering specialization is able to extract the memory stores out of the main computation loop, but uses a smaller loop unroll factor ($2$) than the baseline OpenCL and CUDA versions ($4$ and $16$, respectively).

We were able to obtain speedups of $1.49\times$ and $1.52\times$ over the baseline CUDA and OpenCL versions on the \texttt{GRAMSCHM}, respectively.
Phase ordering is able to move the memory storage operations out of the loop. Aside from that, it uses the same load from memory instruction pattern and unroll factor as the baseline OpenCL version.

The \texttt{MVT} benchmark benefits from phase ordering by a factor of $1.32\times$ and $1.44\times$ over the baseline CUDA and OpenCL versions. The main reason for this improvement is the extraction of the store operation from the computation loop.

The \texttt{SYR2K} benchmarks benefits from phase ordering by a factor of $1.99\times$ and $2.05\times$ over the baseline CUDA version and baseline OpenCL compiled from source, respectively.
In general, the same memory load pattern, loop unroll factor and loop invariant memory storage code motion conclusions apply to this benchmark.
Phase ordering also seems to outline the segment of the code containing the kernel loop, but this does not seem to be the reason for the performance difference.

For the \texttt{SYRK} benchmark, phase ordering improves performance by $1.14\times$ over the OpenCL baselines. 
Once again, the main reason for this improvement is the extraction of the store from the loop.
We could not achieve significant speedups over the CUDA version.

\section{Feature-based Phase Ordering Suggestions} \label{section:features}

It is intuitive to think that the set of compiler passes and compiler sequences that are most suited to compile a given function/program given the target computing platform/device and metrics (e.g., performance, energy, code size) are, at least to some extent, related to static and/or dynamic features of the function/program.
Taking into account such information when deciding what compiler sequences to evaluate during DSE for a given program/function can have the potential to reduce exploration time, or alternatively, achieve better solutions with the same number of compile/execute DSE iterations.

Given the large exploration space for phase selection and ordering, in this paper we propose a simple feature-based approach for suggesting compiler sequences when targeting OpenCL code to GPUs. The experiments show the efficiency of the particular approach and the potential of feature-based approaches in general for compilation targeting GPUs, particularly for use cases demanding fast DSE of compiler phase orders. We believe that the results of these experiments can be of importance in guiding the development of future approaches. 

Given a new OpenCL benchmark, we select $K$ sequences from Table \ref{tab_phase_orders} that are assigned to the $K$ functions/programs most similar with it (k-nearest neighbors).
Those sequences are then used for compilation/evaluation with the new function/program, and the compiled code resulting in the highest performance is selected.
This method allows to suggest compiler sequences very efficiently, and with an overhead that grows in proportion to the value $K$ selected by the compiler user (i.e., number of compilations/evaluations that the user affords).

Phase ordering exploration can have additional termination conditions. For instance, in a real-world use case, instead of terminating only when all sequences from this set having been evaluated or the maximum number of evaluations tolerable by the user being achieved, the DSE could end as soon as the compiled code complies with the non-functional requirements for the function/program.

\subsection{OpenCL code features}

The code-features used in the work presented in this paper are extracted with the MILEPOST GCC toolchain~\cite{fursin:inria-00294704}, which includes the Interactive Compilation Interface (ICI) 2.0 and Machine-Learning (ML) feature extractor version 2.0.
Although the MILEPOST feature extractor was originally developed to extract static C code features, given OpenCL C (the language used to program OpenCL compute kernels) is based on C99, we are using the same flow to extract features from the OpenCL PolyBench/GPU kernels.
MILEPOST features represent an absolute count (e.g., number of basic blocks in the method, number of basic blocks with a single successor) or an average of some count (e.g., average of number of instructions in basic blocks, average of number of phi-nodes at the beginning of a basic block).

Pairs of feature vectors with MILEPOST code features (e.g., feature vectors for function/program from the reference set of functions and the feature vector of a given new function/program) are used to compute a similarity metric that determines the compiler sequences that are suggested for evaluation. 
We did not perform feature selection, thus all $55$ code features extracted by MILEPOST GCC are represented in the feature vectors.
The host code is not taken into account when extracting code features.
Only the OpenCL kernels execute on the GPU, therefore only they are being optimized through compiler sequence specialization.

\subsection{Experiments using code features}

In the experiments presented here we use the cosine distance between feature vectors associated with the OpenCL codes as metric of similarity between pairs of OpenCL codes.
The evaluation of the impact on the results of other similarity metrics is planned for future work.

\begin{figure}[t]
\centering
\includegraphics[scale=0.685]{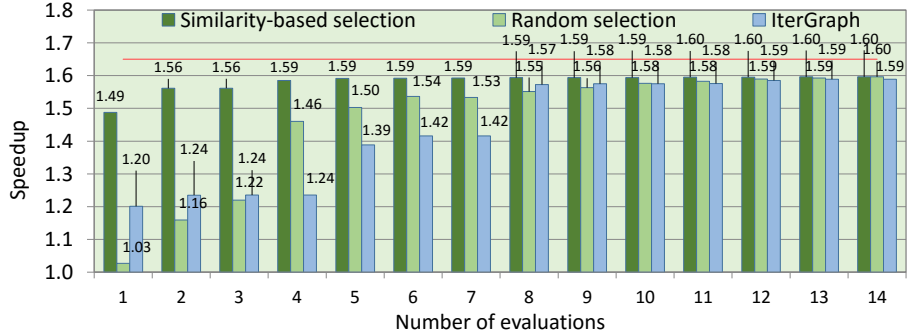}
\caption{Performance improvement on a NVIDIA GP104 GPU for using MILEPOST static code features in the PolyBench OpenCL kernels to select a number of most similar OpenCL kernels and using their sequences. The reference line represents the geometric mean speedup of using the best order for each benchmark, as seen in Figure~\ref{fig:phase_ordering_gtx1070}.}
\label{fig:cosinedistance_vs_random_1_14}
\end{figure}

Figure~\ref{fig:cosinedistance_vs_random_1_14} shows the performance improvement achieved over compilation from source for different numbers of evaluations of compiler sequences from Table~\ref{tab_phase_orders}.
LLVM without any optimization is used as fallback in case no additional sequence evaluated results in better performance.
The approach that relies on evaluating $K$ (a number given by the programmer/user) sequences associated with the $K$ functions/programs more similar with the new OpenCL kernel is compared with the random selection of PolyBench/GPU kernels and using the sequences previously found for them.
Each random selection is performed $1,000$ times, and the geometric mean of the resulting speedups from using the sequences associated with the selected benchmarks is reported.

We also compare with the IterGraph approach (see~\cite{Nobre:2016:GIC:2907950.2907959}), where a graph representing favorable compiler pass subsequences was generated from the sequences from Table~\ref{tab_phase_orders}, using a leave-one-out approach for validation (i.e., when compiling a given OpenCL kernel the sequence associated with that kernel is not used to build the graph).
In fact, all experiments presented in this section are performed with a leave-one-out approach.
When a given PolyBench/GPU kernel is the input of exploration, only the other $14$ PolyBench/GPU kernels and their sequences are considered.

The results show that the selection of the $K$ most similar OpenCL kernels using the cosine distance and using the sequences from Table~\ref{tab_phase_orders} associated with them, results in considerably higher geometric mean performance improvements, for all $K$, when compared with random selection of $K$ OpenCL kernels (and using their sequences from Table \ref{tab_phase_orders}) and also when compared with the use of the IterGraph approach.
With only $2$ additional sequence evaluations, the method based on the cosine distance results in a $1.56\times$ performance improvement compared with compilation w/o phase ordering specialization, while random selection of OpenCL kernels and using the compiler sequences associated with them required $9$ evaluations to achieve the same performance improvement.
For $14$ sequence evaluations, both methods result in the same performance improvement of $1.6\times$, as at that point all sequences from Table~\ref{tab_phase_orders} have been evaluated by both methods.

The IterGraph approach beats random selection of sequences from Table~\ref{tab_phase_orders} for evaluation
when considering up to only $3$ sequence evaluations (on top of LLVM w/o optimization), and after that, only for $8$ and $9$ evaluations.
When considering only up to $14$ sequence evaluations, the IterGraph approach tends to result in compiled code with reduced performance when compared with the simple code-feature based approach presented here.
This may change on new OpenCL kernels from a different benchmark suite or from a different domain.
An advantage of the IterGraph approach in comparison with methods that rely on evaluating predetermined sequences is that new sequences can be generated from the graph (i.e., different from the ones used to build it), which may be more suitable to optimize a given unseen kernel than any of the sequences previously determined to be best for compiling the kernels from a reference set.

The results of our experiments combining the MILEPOST code features extracted from OpenCL C kernels with the use of the cosine similarity as metric for identifying similar OpenCL kernels based on the feature vectors are promising.
Using only the sequence associated with the most similar OpenCL kernel results in a geomean speedup of $1.49\times$ and $1.03\times$, for selection based on cosine similarity and for random selection, respectively.
For $3$ sequence evaluations, the speedups improve to $1.56\times$ and $1.22\times$ respectively.
Finally, for $5$ evaluations, the speedups are $1.59\times$ and $1.5\times$ respectively.
The speedup using $5$ evaluations of compiler sequences suggested using code features is very close to the highest speedup that is achieved when testing the $14$ sequences of all other OpenCL kernels ($1.6\times$).

The performance of the GPU code generated from offline compilation without phase ordering
is close to the performance obtained with online compilation (see Figure ~\ref{fig:phase_ordering_gtx1070}).
For some OpenCL codes, if the number of evaluated sequences is too small, phase ordering can lead to slowdowns over a given baseline (i.e., one of the standard optimization levels, including \texttt{-O0}).
To prevent these situations, it is important to use a safe fallback, such as \texttt{-O3}.
However, when using the similarity-based approach these situations are quite rare.
For instance, considering only the evaluation of a single compiler sequence, \texttt{GESUMV}, \texttt{GRAMCHM} and \texttt{SYR2K} are the only OpenCL benchmarks where the similarity-based selection approach would result in generating code that is worse than the code generated from online compilation.
Two evaluations of sequences are enough to reduce this to only a single benchmark (\texttt{GESUMV}).
In contrast, when randomly selecting sequences to evaluate from Table~\ref{tab_phase_orders}, $8$ OpenCL benchmarks would result in worse performance than online compilation if not using the fallback GPU code for these benchmarks; $2$ evaluations would only reduce that number to $3$.

While the IterGraph approach does not tend to perform very well in comparison with the other two approaches for a small number of evaluations (e.g., up to 14 evaluations in these experiments), it could generate, for any number of sequence evaluations, compiler optimization sequences that result in higher performance for the \texttt{SYR2K} OpenCL kernel than the two other approaches.
We believe that when we validate these approaches using OpenCL kernels from a different benchmark (instead of relying on leave-one-out validation), there will be more cases where the IterGraph approach outperforms the approach for feature-based suggestion of sequences from Table~\ref{tab_phase_orders} evaluated here, and even more cases where it surpasses the efficiency of randomly selecting sequences from Table~\ref{tab_phase_orders}.

Notice that in these experiments, given an OpenCL kernel, the number of compilations and executions of compiled kernels caused by the selection of $K$ kernels from the reference set can be lower than $K$, even for small $K$ values.
Each sequence is only used for compilation once, even if the same sequence is chosen for two or more OpenCL reference codes.
When evaluating the impact of a given selected set of $K$ sequences, subsequent evaluations of identical sequences result in no additional overhead, as the DSE process simply recalls the fitness value (i.e., execution performance) when a previously seen compiler sequence is evaluated for use with the same kernel.
The DSE system also recalls the fitness value for the metric of interest in case a given generated GPU code is identical to the code generated with the use of other compiler sequence previously evaluated, but in this case there is an overhead that incurs because of compilation, which in our experiences is not negligible; especially because we perform validation, during the execution of the main DSE loop, using inputs that result in faster execution time.

\section{Related Work}\label{section:relatedwork}

There have been attempts to identify a small universal set of promising phase orders that could then be used by compiler users to extend the portfolio of the standard optimization levels provided by a compiler (e.g., --O3).
Purini and Jain~\cite{Purini:2013:FGO:2400682.2400715} presented and evaluated approaches for devising a universal set of compiler sequences that is able to cover the program space of a reference set of programs. Given a new program, the sequences from this set are evaluated in a predefined order.
Other authors use features to identify similarities between an unseen code and the codes from a set of reference functions/programs in order to reduce the design space.
For example, the approach presented by Martins et al.~\cite{Martins:2016:CSE:2899032.2883614} relies on special code fingerprints to reduce the number of compiler passes to consider for exploration, reducing DSE overhead considerably; and Amir et al.\cite{amir2016} relies on dynamic code features to guide the generation of specialized compiler phase orders.

The work presented in this paper has key differences to those approaches.
For instance, while other approaches rely on program-features (either static or dynamic) for focus exploration space, our approach uses code-features to select from a set of previously generated sequences.
Unlike sequences generated online using optimization algorithms, these sequences can be strenuously validated by the compiler developers, in a similar way to the validation of the sequences represented by the compiler's default optimization levels (e.g., --O3).
Compiler bugs are a serious hindrance when considering the use of compiler sequence specialization in use-cases where the correct execution of an application is of utmost importance (e.g., safe-critical systems), therefore any approach that drastically reduces the likelihood of the expression of compiler bugs is important.
In addition, the fact that the sequences can be previously validated for correctness allows compiler users to use them with more confidence, reducing the need for extra validation (e.g., testing many pairs of inputs/outputs) than when using the traditional flags such as --O3.
In comparison with the approach as presented by Purini and Jain.~\cite{Purini:2013:FGO:2400682.2400715}, although we also evaluate sequences from a small set of sequences previously generated with iterative compilation, our approach relies on code-features to help efficiently identifying suitable compiler sequences from those sequences.

In this work we identify promising phase orders for a given input function using similarity and code features.
However, an approach such as the one provided by Martins et al.~\cite{Martins:2016:CSE:2899032.2883614}, or any other approach (e.g.,~\cite{Martins:2016:CSE:2899032.2883614,Nobre:2014:ESD:2556863.2556870,Nobre:2016:GIC:2907950.2907959}) that can be used to accelerate iterative DSE can be used to make the initial iterative DSE process required to create the model correlating code-features with suitable compiler sequences faster.

Other authors have also performed a performance comparison between applications written in OpenCL and CUDA. For instance, Fang et al.~\cite{fang} compared benchmarks written in CUDA and OpenCL, and found that the performance differences can be attributed to 4 main factors: programming model differences (e.g., use of texture memory in only one of the versions), different optimizations on the kernels (e.g., one version takes care to coalesce global memory accesses and the other does not), architecture-related differences (e.g., benchmarks that have been tuned for one specific device may perform poorly on others) and compiler/run-time differences. In the benchmarks we tested, we made sure that the code versions are equivalent across the two languages (incl. manual source code optimizations), so the first 3 differences do not apply here. The performance differences can then be explained only by toolchain differences, which explains why our compiler improvements let us bridge the difference between OpenCL and CUDA.
Komatsu et al.~\cite{komatsu} also studied the performance differences between OpenCL and CUDA and found that the CUDA compiler performed significantly more optimizations than the OpenCL code, notably loop unrolling, and that by manually performing the same optimizations, the OpenCL implementation could perform competitively.
Note that in both of these cases, the authors tested the default (\emph{online}) OpenCL implementation, whereas our phase ordering approach is based on an offline compiler.

\subsection{Enumeration-based approaches}

To the best of our knowledge, our work is the first focusing on compiler sequence exploration including phase ordering when targeting GPUs.
However, phase selection and ordering has been focused on by many authors in the context of CPUs, and to a lesser extent FPGAs.
We present next enumeration-based and ML-based approaches for exploring phase orders.

Cooper et al.~\cite{Cooper:1999:ORC:314403.314414} were to the best of our knowledge the first to propose iterative compilation as a means to find phase orders to improve the quality of the compiled code with respect to a given metric.
They used iterative compilation in the form of a GA as a way to minimize the executable footprint.
Cooper et al.~\cite{Cooper2006} explore compiler optimization phase ordering testing different randomized search algorithms based on genetic algorithms, hill climbers and randomized sampling.
Almagor et al.~\cite{Almagor:2004:FEC:997163.997196} rely on genetic algorithms, hill climbers, and greedy constructive algorithms to explore compiler phase ordering at program-level to a simulated SPARC processor.
Huang et al.~\cite{huangfpga} propose insertion-based iterative approaches for compiler optimization phase ordering in the context of hardware compilation targeting an Altera Cyclone II FPGA, to improve circuit area, execution cycles, maximum operating clock frequency, and wall-clock time.
Purini and Jain~\cite{Purini:2013:FGO:2400682.2400715} propose an approach that relies on a list of compiler sequences previously found to be suitable, relying on iterative algorithms, for a reference set of benchmarks representative of all classes of programs.
Their approach circumvents program classification by relying on a small set of sequences which has the particularity of including an optimization sequence for each possible program class.
Given a new program, each of these compiler sequences is tested and the one leading to better binary execution performance after compilation of the new program is selected.
Nobre~\cite{nobrefpl2013} presents results for the use of a SA-based approach to specialize compiler sequences in the context of software and hardware compilation.
More recently, Nobre et al.~\cite{Nobre:2016:GIC:2907950.2907959} presented an approach based on sampling over a graph representing transitions between compiler passes, targeting the LEON3 microarchitecture.

\subsection{ML-based approaches}

Agakov et al.~\cite{agakov2006} present a methodology to reduce the number of evaluations of the program being compiled with iterative approaches.
Models are generated taking into account program features and the shapes of compiler sequence spaces generated from iteratively evaluating a reference set of programs.
These models are used to focus the iterative exploration for a new program, targeting the TI C6713 and AMD Au1500 embedded processors.
Kulkarni and Cavazos~\cite{Kulkarni:2012:MCO:2384616.2384628} proposed an approach that formulates the phase ordering challenge as a Markov process where the current state of a function being optimized conforms to the Markov property (i.e., the current state must have all the information to decide what to do next).
Instead of suggesting complete compiler sequences, these authors use a neural network to propose the next compiler pass based on current code features.
Sher et al.~\cite{Sher:2014:PRN:2568326.2568328} describe a compilation system that relies on evolutionary neural networks for phase ordering.
Neural networks constructed with reinforcement learning output a set of probabilities of use for each compiler pass, which is then sampled to generate compiler sequences based on the input program/function features.
Martins et al.~\cite{Martins:2016:CSE:2899032.2883614} propose the use of a clustering method to reduce the exploration space in the context of compiler pass phase order exploration.
Amir et al.~\cite{amir2016} present an for compiler phase ordering that relies on predictive modeling, using dynamic features to suggest the next compiler phase to execute to maximize execution performance given the current status.

\section{Conclusions}\label{section-conclusions}

This paper showed that compiler pass phase ordering specialization allows achieving considerable performance improvements when compiling OpenCL kernels to GPUs.
In our first set of experiments we explored the performance impact of specialized phase orders obtained by iterative compilation.
In our second set of experiments we explored the use of a feature-based approach to identify specialized phase orders.

Targeting an NVIDIA GP104 GPU using Clang, we were able to improve the performance of code compiled from PolyBench/GPU OpenCL kernels  $1.65\times$ on average (up to $5.70\times$) over the default compilation flow. 
The use of phase ordering on top of the OpenCL versions of the kernels resulted in a geometric mean speedup of $1.54\times$ (up to $5.48\times$) when compared with the performance of the equivalent CUDA kernels compiled with NVCC.
We gave insights explaining why using specialized phase orders tends to result in speedups. We found that due to phase ordering, the compiler was able to extract memory writes from loops by using an accumulator register, reducing the number of expensive global memory writes.

We presented results that give confidence that static features in OpenCL kernels can be used in the context of suggesting suitable compiler sequences more efficiently when targeting GPUs. 
Evaluating sequences, previously found in an initial exploration phase to be more suitable for compiling OpenCL code from a reference set, ordered by similarity to the new unseen OpenCL code, results in higher optimization with the same number of compiler sequence evaluations on an NVIDIA GPU; compared both with randomly evaluating sequences from this set of sequences and with evaluating sequences generated by the IterGraph approach.
Using a leave-one-out validation approach, geometric mean performance improvements of $1.49\times$, $1.56\times$ and $1.59\times$, versus the same OpenCL baseline, were achieved while evaluating only $1$, $3$ and $5$ compiler sequences from this set, respectively. For comparison, our best phase orders found by our initial experiments lead to improvements of $1.65\times$.

We are currently evaluating how to better extend our compiler phase ordering exploration framework to allow exploration targeting the CUDA kernels instead of only OpenCL kernels.
We are also evaluating the potential of compiler phase ordering for GPU energy consumption reduction, and accessing how it correlates with execution performance, as GPUs are used in domains with energy (and power) concerns (e.g., HPC, embedded), so there may exist scenarios were it is acceptable to sacrifice performance for less total energy use.

As future work, we plan to analyze experimental data about phase orders that result in problematic or incorrect situations, to create a model to evaluate the likelihood that a new compiler sequence will be unsuitable and reduce the number of evaluations for these sequences.

\subsubsection*{Acknowledgments}

This work was partially supported by the TEC4Growth project, NORTE-01-0145-FEDER-000020, financed by the North Portugal Regional Operational Programme (NORTE 2020), under the PORTUGAL 2020 Partnership Agreement, and through the European Regional Development Fund (ERDF). Reis acknowledges the support by Funda\c{c}\~{a}o para a Ci\^{e}ncia e a Tecnologia (FCT) through PD/BD/105804/2014.
In addition, we acknowledge Tiago Carvalho for his work on the LARA framework, used to develop the DSE tool used for this paper.

\bibliographystyle{splncs04}
%\bibliography{biblio}

\end{document}